\documentclass{sig-alternate-05-2015}

\def\etal{{\it et al.}}

\usepackage{graphicx}
\usepackage{array}

\usepackage{multirow}
\usepackage{epstopdf}
\usepackage{color}
\usepackage{booktabs}

\title{Towards Bottom-Up Analysis of Social Food}

\begin{document}

\CopyrightYear{2016} 
\setcopyright{acmlicensed}
\conferenceinfo{DH '16,}{April 11 - 13, 2016, Montréal, QC, Canada}
\isbn{978-1-4503-4224-7/16/04}\acmPrice{\$15.00}
\doi{http://dx.doi.org/10.1145/2896338.2897734}
\clubpenalty=10000 
\widowpenalty = 10000 

\numberofauthors{1} 
\author{
\alignauthor
Jaclyn Rich, Hamed Haddadi, Timothy M. Hospedales\\
\affaddr{School of Electronic Engineering and Computer Science}\\
\affaddr{Queen Mary University of London}\\
}

\maketitle
\begin{abstract}

Social media provide a wealth of information for research into public health by providing a rich mix of personal data, location, hashtags, and social network information. Among these, Instagram has been recently the subject of many computational social science studies. However despite Instagram's focus on image sharing, most studies have exclusively focused on the hashtag and social network structure. In this paper we perform the first large scale content analysis of Instagram posts, addressing both the image and the associated hashtags, aiming to understand the content of partially-labelled images taken \emph{in-the-wild} and the relationship with hashtags that individuals use as noisy labels. In particular, we explore the possibility of learning to recognise food image content in a data driven way, discovering both the categories of food, and how to recognise them, purely from social network data. Notably, we demonstrate that our approach to food recognition can often achieve accuracies greater than 70\% in recognising popular food-related image categories, despite using no manual annotation. We highlight the current capabilities and future challenges and opportunities for such data-driven analysis of image content and the relation to hashtags. 

\end{abstract}




\section{Introduction}

Social networking has become a mainstay of modern life; 79\% of all online adults use at least one social media site~\cite{Duggan2015}. Usage of Instagram, especially, has become increasingly widespread. Instagram is a social networking mobile phone application that allows users to upload and share photos and videos. Using hashtags to describe one's posts allows the user to attach sentimental or contextual information to the pictures that appear on their timeline; Instagram permits a maximum of 30 hashtags to be attributed to a post. Instagram also allows the user to geo-tag, or assign a geographical location, or point of interest to, their individual images. Since its inception in October 2010, Instagram has grown to over 500 million monthly users, and an average of 70 million photos uploaded daily; the rise in Instagram's popularity can be partially attributed to the rise in the usage of smartphones and the high quality cameras contained within them. 

Food is an important part of all cultures. As individuals social interactions have moved more towards being digital, their way of communicating about food has too. Documenting food intake has become a phenomenon, to the point of inspiring parody. One such lampoon, ``Eat It Don't Tweet It (The Food Porn Anthem)" contains the lyrics, \emph{``its unthinkable to dine out and not record it"}, which does not strike too far from the truth for many social media users.\footnote{Eat It Don't Tweet It (Instagram Food Porn Anthem) \url{https://www.youtube.com/watch?t=93&v=WpNIt6UC8uo} [Video file last accessed November 2015] } Motivations for documenting one's consumption on social media include the desire to share their healthy habits and their indulgences.\footnote{10 Reasons Why People Post Food Pictures on Facebook, available on \url{https://www.psychologytoday.com} [retrieved 31 July 2015]} Analysis of food images on Instagram can lead to further understanding of food from health, cultural, or marketing perspectives. Improving the world's health is of the utmost importance, as 30\% of the global population is overweight or obese. Urban areas are those that are seeing the greatest increase in obesity. These regions are also those that have the highest usage of Instagram, making them both the most in need of and accessible for intervention~\cite{hu2014we}.

Understanding nutritional content of our food is integral to good health. Many health programs take this a step further by requiring its participants to track and possibly record the nutritional information of the food and drink they ingest. This is often done by manually entering each food item. Consequently, an accurate and simple way of counting calories is important to guide and inform individuals food choices.  Additionally, work by Zepeda and Deal~\cite{zepeda2008think} found that photographic food diaries were more effective than written ones at prompting patients to truly understand their eating habits and spur a change in their behaviour. Keeping a visual food diary that also instantly provides nutritional content could make for a powerful approach to weight-loss. Furthermore, Instagram food images will provide a more accurate portrayal of people's eating habits than self-reporting. When self-reporting, teenagers tended overestimate to their intake, while adults tended to underestimate~\cite{black1993measurements}.  

Creating a means of recognising foods is one of the first steps in creating accurate automated calorie intake counting applications, e.g., using pictures of meals taken by an individual during their day. Previous works have mostly focused on use of hashtags and keywords for understanding calorie value of image contents. Using classifiers trained bottom-up on images labelled with hashtags would be both more scalable, and less biased than having to manually define food ontologies and label the content of images.

In this paper we perform a joint analysis of Instagram hashtags and images, in order to: (i) reveal the limitations of conventional non-visual tag-centric analyses of social media, and (ii) explore to what extent these can be addressed by further analysing image content. We look at Instagram image content using visual recognition techniques. Differently to the conventional approach of defining a food category ontology and curating an associated database to train visual recognisers; we explore the possibility of hashtags themselves being considered ``visual".  This was accomplished through the creation of a series of hashtag classifiers to ascertain how visually recognisable the user-generated tags are. We contrast the results obtained with those using ImageNet~\cite{deng2009imagenet}\footnote{\url{http://www.image-net.org/}} as the object ontology to recognise. Moreover, we investigated the cultural differences in documented food across the globe by analysing the data from a spatial perspective. The correlations between different visual tags were also determined. We also extensively discuss the practical challenges in accomplishing these tasks including computational those induced by label-noise. 

\section{Related Work}
\label{Sec:related}

\subsection{Food in Social Media}

Fried~\etal~\cite{fried2014analyzing} demonstrated ``the predictive power of the language of food" through the analysis of a large corpus of tweets. They only retrieved tweets that contained meal-related hashtags: dinner, breakfast, lunch, brunch, snack, meal, and supper.  They further filtered these tweets to find mentions of food and used the term frequency-inverse document frequency (TF-IDF) of these mentions to recognise uniquely characteristic foods for each state. While some of these representative foods confirmed already held understandings of state food preferences, such as Texas' affinity for brisket and the South's for grits, some like North Dakota's association with flan were more surprising.

Abbar~\etal~\cite{abbar2014you} explored the relationship between the calories in the food tweeted by users in a state with the state's obesity and diabetes rate. They discovered a Pearson correlation of 0.772 with obesity and 0.658 with diabetes. On average, women and residents of urban areas tweet about food with fewer calories than men and their rural counterparts. Tweets from urban areas are also characterised by greater mentions of alcoholic beverages and such foods as avocado and crab. In contrast, tweets from rural areas are concerned with more conventional fare, such as pizza, chocolate, and ice cream. 

\vspace{1cm}
\subsection{Instagram users' behaviour}

Instagram is a relatively young social media site, and hence has not attracted as much research as its more established counterparts. Primary research focused broadly on user practices. Hu~\etal~\cite{hu2014we} performed one of the first studies using Instagram data and identified eight image content categories, with the first six being the most popular: selfies, friends, activities, captioned photos (pictures with embedded text), food, gadgets, fashion, and pets. 

Using keyword-based calorie lookup tables, Sharma and De Choudhury~\cite{Sharma:2015:MCN:2740908.2742754} analysed a dataset of Instagram posts that contained food-related tags.  They were able extract nutritional content for 93.5\% of Instagram posts in their dataset by associating each hashtag with a food item and retrieving its nutritional information in the USDA National Nutrient Database.  Using crowdsourced verification, they were found to be 89\% accurate in their calorie estimations.  They used these findings in conjunction with feedback from Instagram users, and found that moderately caloric foods (150 - 300) attracted the most likes and comments. The authors recognised the value of future works incorporating image analysis into this area of study.

Mejova~\etal~\cite{mejova2015foodporn} analysed a large dataset of food images posted on Instagram from Foursquare restaurant locations. They identified the most popular foods that spurred the use of the \textit{\#foodporn} hashtag across the United States.  They found that images of Asian cuisine were the most likely to be tagged with \#foodporn; cuisine from slow food restaurants, as compared with fast food establishments, was also more likely to be tagged as such. In the health arena, they first investigated the correlation between the check-ins at fast food restaurants on Foursquare in a county, to the percentage of the county population that is obese; they found a Pearson correlation of 0.424.  Hashtags that had high prominence in counties with high obesity, such as \#ilovesharingfood, and low obesity, such as \#smallbiz, were also identified.

\subsection{Image recognition on food}

Research and industry have started using Machine Learning (and specifically Deep Learning) for object and scene recognition tasks.\footnote{\url{http://yahoohadoop.tumblr.com/post/129872361846/large-scale-distributed-deep-learning-on-hadoop?soc_src=mail&soc_trk=ma}} Kawano and Yanai~\cite{Kawano:2014:FIR:2638728.2641339} developed classifiers to recognise types of food from images using various features. Their experiments revealed that using Deep Convolutional Neural Networks (DCNN) along with RootHoG patches and colour patches coded into a Fisher Vector representation had the best performance of 72.26\% classification accuracy.  When used separately, the DCCN outperformed all of the other features. They further developed a mobile application for Android called FoodCam that recognises food items from images and provides nutritional information in real-time. They achieved 79.2\% classification accuracy for the top five category candidates.  Although they previously explored using DCNN, here they used a histogram of oriented gradients patch descriptor and a colour patch descriptor with a Fisher Vector representation so that it could run on a mobile phone. Linear SVMs were used as the classifier. The system asks the user to draw a bounding box around the food, and proposes the five most likely foods that it could be from which the user can select one. Nevertheless, this more conventional approach has key limitations: It requires pre-specification of the food categories to be detected, and manual collection and annotation of images of these categories in order to train detection models. This means that (i) It is not scalable to very many categories of food due to the barrier of manually collecting and annotating images of each and (ii) Such a top-down driven approach only recognises the pre-specified food ontology, which may not completely overlap the particular categories that people pervasively eat and share on social media.

In related research, CNNs are used for detection of specific classes of fruits and vegetables from ImageNet, with an accuracy of 45-75\% for the top categories~\cite{morteza2016}. The results were reported to be highly affected by the quality of the images and the number of objects in each image. This work was limited to carefully labeled images of specific categories. In this paper, we use Instagram pictures in order to assess the ability to perform such analysis on pictures taken by individuals in natural settings.
 

\subsection{Limitations}
While these studies provide useful insights about people's food consumption habits, they are limited in their scope. Data from Foursquare and Yelp are limited in that they only contain information and images of food or drinks prepared at restaurants. The food images from Instagram can include home cooked meals, packaged foods, ingredients, etc.  In addition, Foursquare's community of 55 million is also much smaller than Instagram's 400 million. There is also much work to be done to create accurate calorie count databases from images. In Sharma and De Choudhury's work, they claimed 89\% accuracy in predicting calorie content.  Yet in one of the analysed posts, the mentioning of the ``Ultimate Red Velvet Cheesecake" from The Cheesecake Factory, was simplified to cheesecake.  While on average a slice of cheesecake may contain the predicted 402 calories, according to MyFitnessPal, a slice of this specific cheesecake has 1,250 (``Calories in the Cheesecake"). This further illustrates how the rich diversity of food categories worth recognising easily goes beyond common pre-specified top-down ontologies as highlighted earlier.

\section{A Case Study on Food}

In this paper, we analyse food images takes on Instagram for content. We use a large corpus of data from Instagram, in addition to labeled data form ImageNet for testing and training our classifiers.

\begin{table}[t]
\centering
\caption{Most Frequent Hashtags in the Dataset.}\label{tab:hashtagRank}  
\begin{tabular}{ll}
\hline
Rank & Hashtag   \\
\hline
1 &	food \\
2	& foodporn \\
3	 &  instafood\\
4 &	yummy\\
5 &	delicious\\
6 &	foodie\\
7	& instagood\\
8 &	yum\\
9 &	foodgasm\\
10 &	dinner        \\   
\hline
\end{tabular}
\end{table}

\vspace{0.5cm}
\subsection{Dataset}
We used the Instagram API\footnote{\url{https://instagram.com/developer/}} to download images with food-related tags on during two different periods, autumn/winter of 2014 (first and fourth week of October, all four weeks of November, first week of December) and spring of 2015 (third and fourth week of March). We also collect the associated metadata, such as the number of likes, comments, captions, GPS location, and hashtags. The data set is comprised of a total of 808,964 images and their corresponding metadata. We crawled the images and data for the 43 most popular food-related hashtags as identified by Mejova~\etal~\cite{mejova2015foodporn}. These tags include the likes of \textit{\#foodporn, \#yum, \#foodie, \#instafood, \#lunch, \#delicious, \#foodgasm}, etc. We also crawled the the 53 most popular hashtags of specific food items which include tags like \textit{\#fpizza, \#rice, \#muffin, \#pasta, \#chicken, \#donut, \#steak, \#sushi, \#kebab}. The ranking of most popular hashtags in the collected dataset is presented in Table~\ref{tab:hashtagRank}. We did not apply any pre-filtering to our data in order to avoid introducing biases.\footnote{We are making our image dataset available with this paper via \url{http://www.eecs.qmul.ac.uk/~tmh/downloads.html}}

\subsection{Parsing of the Image Metadata}

Analysing these images with state of the art computer vision techniques, we wish to answer the following questions:

\begin{itemize}
\item How many of these images are food-related at all?
\item How many pictures have images that are indeed relevant to their hashtag?
\item Of the tags associated with these images, how many and which of them are \emph{visual}; in the sense that they correspond to a visible object (such as plate), rather than a non-visual concept (such as a good night out).
\item And what can visual analysis of Instagram posts tell us beyond standard tag-based analyses.
\end{itemize}

\subsection{Feature Extraction}

A total of three different features were extracted during the preliminary experimentation phase. First, the GIST descriptor, having 512 dimensions, was extracted.  The GIST feature is a low dimensional representation of an image region and has been shown to achieve good performance for the scene recognition task when applied to an entire image~\cite{Oliva:2001:MSS:598425.598462}. Classification using this low dimensional descriptor yielded results that were not at all or not much better than random.  This led to the decision to use a higher dimensional feature.

The second and third features used are from pre-trained convolutional neural networks (CNNs). CNNs were chosen since they achieve state of the art results for general image category recognition \cite{simonyan2015veryDeep}. The CNNs used were implemented using the MatConvNet toolbox.\footnote{\url{http://www.vlfeat.org/matconvnet/}} These are namely \textit{imagenet-vgg-f} and \textit{imagenet-vgg-verydeep-16}, the latter a CNN with 16 layers~\cite{simonyan2015veryDeep}. The 16-layer very deep model was selected for our work as it had higher precision than the other in preliminary experiments.  

The CNN provides two capabilities, and we use both: (i) Its final layer provides a probability of how likely it is to be belong to each of the 1,000 classes in ImageNet \cite{deng2009imagenet}, as well as a final classification into the most likely class. ImageNet is a database of images built upon the nouns of WordNet, a lexical database of the English language.  It contains over 500 images for each of over 1,000 categories. (ii) Its  penultimate layer also provides a 4,096 dimensional feature vector, which serves as a powerful image representation for learning to recognise new classes outside of those covered by ImageNet. 

\subsection{Classification Model}
The hashtag classification was performed using the Support Vector Machines (SVM). During our initial exploratory experimentation phase, both linear and radial basis function kernels for SVM were applied.  The linear classifier performed better (up to 5\%) in all trials. For this reason and due to its much lower computation cost, the linear kernel was chosen for experimentation using the entire dataset. In our main experiments, we used the liblinear~\cite{fan2008liblinear} primal SVM optimiser due to its suitability for the large scale data explored here.

The principle behind SVM classification is that the optimal decision boundary is the one that maximises the margin between the boundary and the data points that are closest to it (the support vectors) and minimises misclassification errors. Additionally, during the small-scale experiments, 10-fold cross validation was used to set the cost parameter. This greatly increased computation time, and therefore was not used during the final training phase over the entire dataset. In future iterations in this space, this approach could be employed to attempt to improve upon the results found in this paper. 

\subsection{Reverse Geocoding}

\begin{figure}[t]
\centering
\includegraphics[width=1.0\columnwidth]{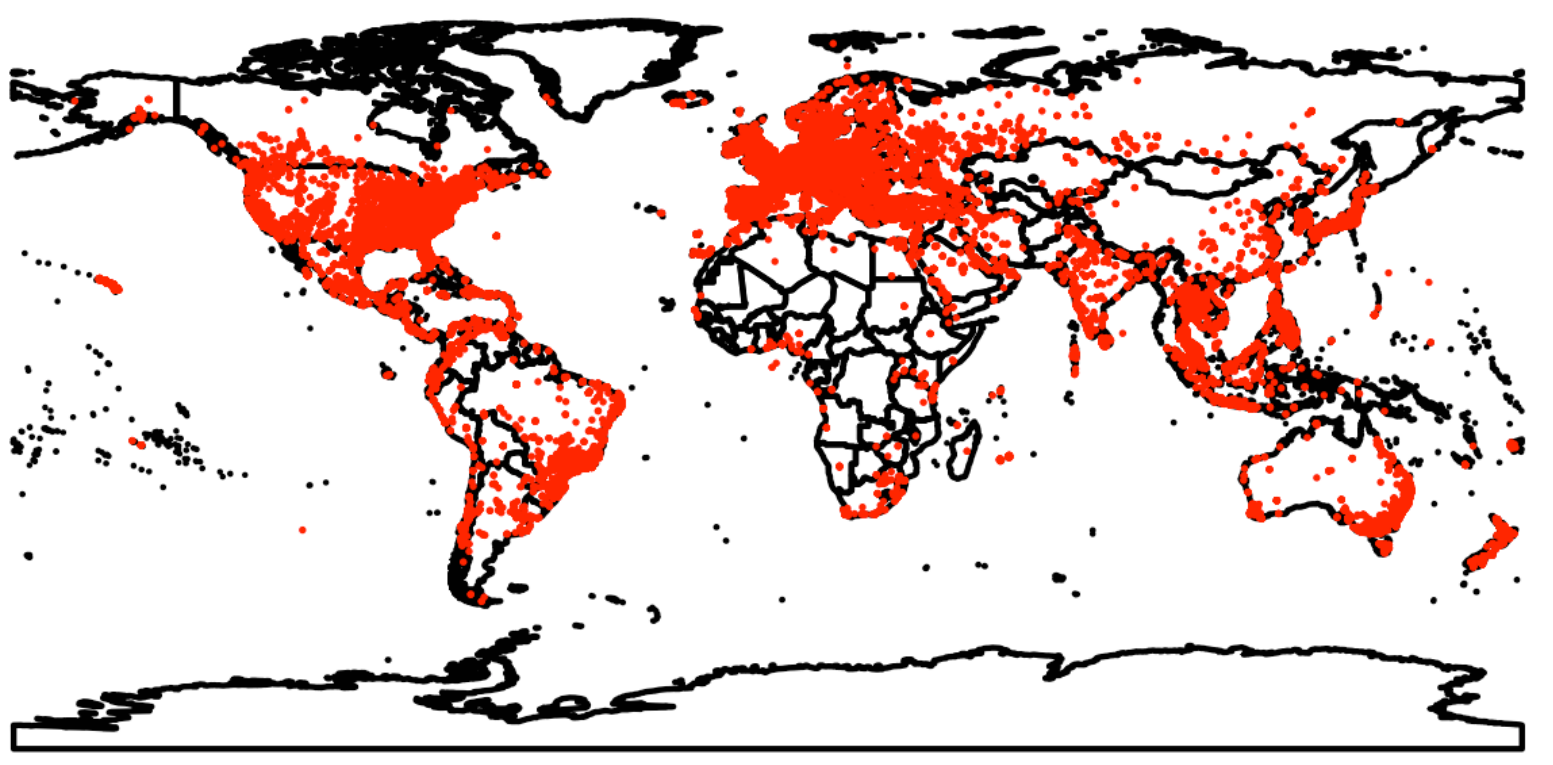}
\caption{Map of all geotagged posts in the dataset\label{fig:geotag}}
\end{figure}

Reverse geocoding is the process of converting longitude and latitude coordinates to an address. Our interest was in obtaining the country and continent information only. A post may contain an Instagram location that corresponds to a location on Foursquare, latitude and longitude, or no location data.  Once location data was extracted from each metadata structure, the \textit{sp}, \textit{maps}, \textit{rworldmap}, and \textit{rworldxtra} packages in R were used to project the GPS coordinates onto a map and report the country and continent of the projection. The country and continent were recorded in the metadata structure for each entry with location data.  While the country and continent were obtained successfully for most of those posts with location data, costal locations were frequently not identified as they were frequently mistaken for lying in bodies of water. The GoogleMaps API was then used on these unidentified costal points to minimise the number of points for which an address could not be retrieved.  Out of 808,964 posts in our dataset, 315,806 of them (i.e.,39\%) were geotagged. Out of the total of these 315,806 geotagged posts, 301,234 were successfully reverse geocoded. Majority (\~30~\%) of the geocoded posts were, in decreasing order, from Asia, Europe, and North America. In terms of countries, United States, Australia, Italy, and the UK had the most frequent geotagged posts. Figure~\ref{fig:geotag} displays the map of our geotagged posts.

\section{How Noisy are Instagram Food Tags?}
\label{sec:TagQual}

Since Instagram Tags are user generated, there is the question of to what extent the food-related tags we have crawled, and are interested in studying, actually correspond to food-images? Some tags might appear to refer to foods, but actually represent Internet vernacular, and thus the image could represent anything. Other tags might refer to foods, but the associated image reflect some other aspect of the dining experience such as menu or people. Social media posts are often used for understanding food consumption and culinary interests across the world and correlations with obesity and other health conditions. The reasons we want to know about tag-image content correspondence are two-fold: (i) Because previous work has used food-related tags or keywords as an indicator of actual food consumed \cite{mejova2015foodporn}. Quantifying the prevalence of food-tags on non-food images may indicate something about the expected noise level in such studies. (ii) Because it will shed some light on the amount of noise in tag-image correspondence, and hence how well we can expect to learn from the tags later on.

\vspace{0.1cm}\noindent\textbf{What Proportion of Food-Tagged Images Actually Contain Food?}\quad 
Since evaluating the accuracy of user`s tagging manually is implausible for a dataset of this size, we resort to automated approximation. In the first experiment, we address this question in a coarse way by finding out the proportion of food-tagged Instagram images (all of our dataset) that actually correspond to (any kind of) food, and the proportion that do not correspond to any kind of fond. The state of the art CNN (VGG~\cite{simonyan2015veryDeep}) we use to analyse images is trained on the ImageNet library which covers 1,000 classes, 61 of which are food related. Based on the CNN output, we can estimate upper and lower bounds of the amount of actual food and non-food images in the dataset. To estimate the lower-bound on true food images, we count those images confidently\footnote{Defined as above 0.5 posterior -- a conservative threshold since the posterior is normalised over 1,000 categories.} classified as one of the food types, and also those images confidently classified as one of the food containers (6 categories including plate, bowl, etc) that ImageNet recognises, since the classifier often recognises the container more confidently than the food. To estimate the upper bound on true food images, we count those images confidently classified as one of the 933 (1000-61-6) categories which are neither food nor food-container related.

\begin{figure}[t]
\centering
\includegraphics[width=1.0\columnwidth]{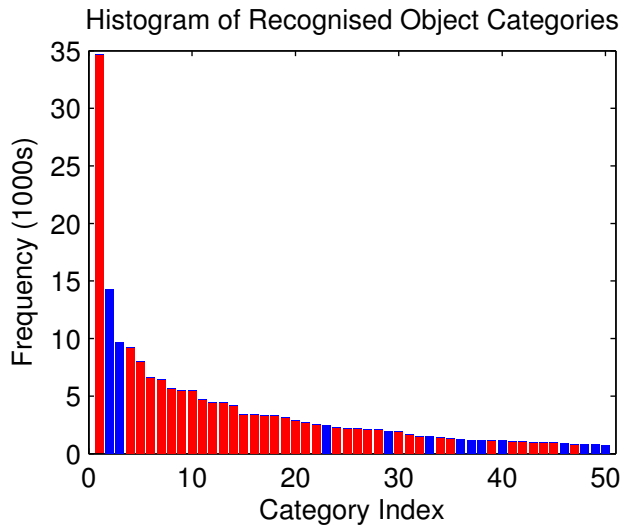}
\caption{Food (red) and non-food (blue) dataset category distribution (top 50), as recognised by VGGnet \cite{simonyan2015veryDeep}\label{fig:foodNonFoodDist}}
\end{figure}

\begin{figure}[t]
\includegraphics[width=1.0\columnwidth]{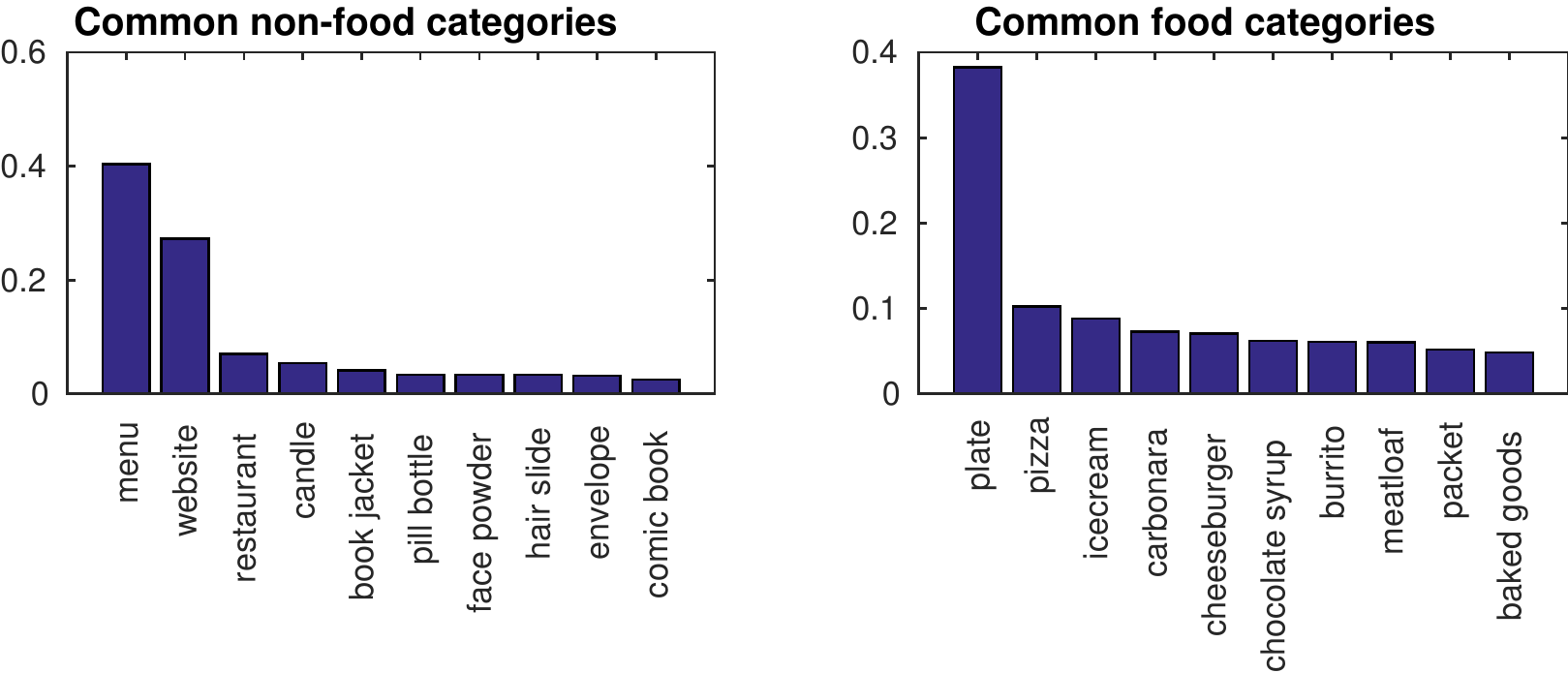}
\caption{Commonly found food-related (right) and non-food-related (left) categories among VGGnet \cite{simonyan2015veryDeep} confidently recognised categories \label{fig:vggCatsFound}}
\end{figure}

Figure~\ref{fig:foodNonFoodDist} shows the distribution of the top-50 most common and confidently recognised categories, coloured according to food-related (red) and non-food-related (blue). By this measure, the total proportion of food images to all is around 19\%, and the proportion of non-food images is 11\%. These figures thus provide a rough lower and upper bound (19-89\%) for the amount of actually food-related data on Instagram when crawling food tags. Figure~\ref{fig:vggCatsFound} illustrates the most commonly found food and non-food categories in the dataset. Websites and menus are by far the most common non-food images that are presented with food-related tags. It is easy to understand the reasons behind users posting menu photos with food-related tags, though the reason for a high number of website screenshots is not immediately clear.

Naturally these results are contingent on the reliability of the VGG CNN recogniser. However, the majority of the highly confidently classified images used to generate the above are correctly classified, when checked manually. By lowering the confidence threshold used, we could make the bound tighter, but as the VGG recognisers would be making more errors at this point, we would also be less confident about its correctness at each end.

\section{Exploiting bottom-up vision}\label{sec:visualHashtagResult}

In the previous set of experiments, we used the state of the art VGG-CNN to recognise categories in InstagramPosts. This followed the conventional pipeline \cite{Kawano:2014:FIR:2638728.2641339} of pre-specifying an ontology and obtaining curated and annotated data (e.g., ImageNet), and using this data to train a recogniser for categories in this ontology (e.g., VGG-CNN).  As outlined earlier, that approach is limited in scalability (need to collect and annotate data for every category to recognise), and top-down bias of ontology selection. In this section, we explore an alternative approach of discovering visual concepts automatically tags, thus avoiding the scalability barrier and ontology selection bias intrinsic to earlier work.

\subsection{Discovering Visual Food Tags}

The first question we ask is which tags correspond to visual concepts (like hamburger, plate, table) -- and thus provide a potential route to automatically learning to recognise foods or other content in images; and which tags correspond to (presumably) abstract non-visual tags such as user emotions  (\#hygge), or abstract concepts (\#instagood), etc. To investigate this we take the approach of going through each tag and attempting to train a classifier for it. This approach has been used before \cite{berg2010automatic_web,layne2014wildAttr,chen2013neil}, but never at this scale or focus on food. The score of each classifier on a test split of the data reflects the tags visualness; i.e., there may be no visual regularity in photos tagged with \#instagood, hence the test score will be low. But there may be high visual correlation between the tag \#salad and associated photos so the score may be higher. Specifically, we take the top 1,000 most common tags and train a classifier for each. This is a large scale operation that took about 2 Xenon E5 CPU-weeks.

\begin{figure}[t]
\centering
\includegraphics[width=0.9\columnwidth]{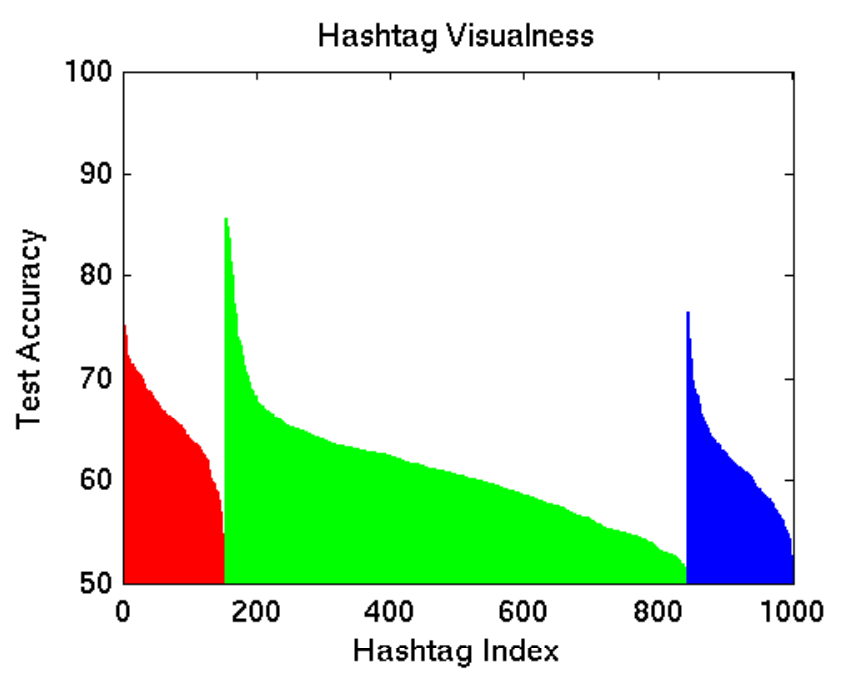}
\caption{Visualness of the top-1000 hashtags. Sorted by our manual expected categorisation: red -- concrete food, green -- non-food, blue -- food-related.}\label{fig:visualness}
\end{figure}

\vspace{0.1cm}\noindent\textbf{Which Tags have Visual Correlates?}\quad Before focusing on food-specific results, we explore the overall visualness results across the entire top-1000 hashtag dictionary. To break down these results, we manually classified each of the top 1,000 hashtags according to whether the associated concept was: (i) A concrete food object: a noun representing a physical food object that can be pictured, such as hamburger/salad/cappuccino. Once detected, such concrete objects can provide the input to calorie counting~\cite{Kawano:2014:FIR:2638728.2641339,mejova2015foodporn,Sharma:2015:MCN:2740908.2742754}, etc. (ii) More abstract food-related terms that are not specific objects, such as categories (breakfast/dinner or chinesefood/koreanfood) or attributes (healthy/highcarbs). Although not specific food objects, these may still be of interest for digital health researchers. (iii) Non-food related terms. We expected that the concrete food objects would be potentially `visual' (consistently recognisable), while it was unclear to what extent the abstract food-related terms would be visual. 

Figure~\ref{fig:visualness} summarises the empirically evaluated visualness of each tag, broken down by our manual categorisation. From this we observe that: (i) The majority of tags (691 tags, green) are actually not food related at all, indicating a high-level of noise in Instagram data (either unrelated photos picked up by our crawler due to having spurious food related tags, or food photos with non-food-related tags). (ii) Most of the concrete food object tags (152 tags, red) are quite visual, with 133 of them having test accuracies in the range of 60-77\%. (iii) Of the abstract food-related tags (157 tags, blue), a significant fraction are indeed quite visual, with 98 of them having test accuracies in the 60-77\% range.

Overall, the top most visual hashtags were \#sneakers, \#pants, \#polo, \#jeans  \#tshirt, all of which we had categorised non-food related (Figure~\ref{fig:visualness}, green). The most visual hashtags in our food-related but abstract category (Fig.~\ref{fig:visualness}, blue) included a variety of tag types: Food containers such as \#lunchbox are understandably visual. Food ethnicities such as \#koreanfood, \#chinesefood, \#japanesefood, \#vietnamese, were highly visual, implying that Asian foods are relatively easy to identify visually (e.g., in contrast \#italianfood and \#frenchfood were ranked low for visualness). Some comments on food type such as \#healthybreakfast and \#homecooked were also visual, which is somewhat surprising as it is not obvious that home cooked food should be visually discernible. Finally \#foodporno and \#foodpornasia were also highly visual suggesting that, although abstract, the types of foods that \#foodporno is applied to are sufficiently consistent for images warranting this tag to be detectable.

\vspace{0.1cm}\noindent\textbf{Which Food Tags are Visually Recognisable?}\quad We next focus more specifically at the concrete food object tags (Fig.~\ref{fig:visualness}, red) for quantitative analysis. The results for the top-20 most visual food hashtags are shown in Table~\ref{tab:foodVisual}. These hashtags were determined as \emph{visual} by the metric of validation accuracy. Going beyond the automated visualness metric of cross-validation accuracy, we also manually checked the reliability of these results by inspecting the top-50 most highly ranked test images for each hashtag and labelling the true and false examples. From this we could compute Precision @ 50, and average precision (over all ranks to 50), also shown in the table. It shows that in general, confidently learned (high visualness) tags do indeed result in reasonably high precision food recognisers, with some exceptions. This noteworthy as, in strong contrast to the conventional pipeline, these are bottom-up results without top-down pre-specification of a food ontology, and costly no manual curation and annotation of training data. This suggests that our approach could be a more scalable route to ultimately recognising a very wide variety of food categories in future.

\subsection{Improving Food Image Recognition}\label{sec:improve}

\vspace{0.1cm}\noindent\textbf{Ameliorating Label Noise}\quad The reliability of the previous results are impacted by a variety of factors: (i) false positives such as \#dinner being used to refer a variety of situations around dining (such as friends) rather than the meal itself, and (ii) false negatives such as tagging \#foodporn but not \#cake on a cake photo results in a classifier being trained to treat that cake photo as an example of a non-cake image. We attempted the former of these issues by developing a filtering strategy of recognising frequent and reliable VGG classes as distractors, and removing them from the dataset to prune false-positive label-noise. In particular we removed images recognised as \emph{website, restaurant, book jacket, comic book} and \emph{wig} from the dataset. We verified that all these classes have at least 95\% precision @ 20, so we are confident that the label noise is being removed. Restaurant images contain restaurants with no food; detected wigs are typically female selfies with long colourful hair, etc. This pruned the dataset of 48,031 distractor images that were definitely not food. However, repeating the previous visual tag discovery experiment, the analysis showed only a small improvement on the resulting visualness and manually verified precision compared to Table~\ref{tab:foodVisual} so we omit the results here.

\vspace{0.1cm}\noindent\textbf{Focusing on Plates}\quad In order to find a stronger approach to improving label noise, and to provide a slightly different view of the results, we next developed a filtering strategy based on focusing on food containers. Specifically, we exploited VGGnet ability to reliably recognise the \emph{Plate} class. We then selected only those 105,836 images that contained confidently recognised plates to repeat the analyses in Section~\ref{sec:visualHashtagResult}. Although selecting only plate images excludes a large volume of food data, it also removes almost all of the widely prevalent non-food, distractor images (Figure~\ref{fig:vggCatsFound}, left), since plate photos almost always contained food.

\vspace{0.1cm}\noindent\textbf{Results}\quad From the updated results in Table~\ref{tab:foodVisualPlate}, we see quite a different picture compared to the un-filtered results in Tab.~\ref{tab:foodVisual}. In particular, we make the following observations: (i) The top 20 hashtags are all food-related, unlike the previous experiment where food-related hashtags were ranked 10-47th. (ii) Some Japanese foods are now the most visual hashtags, presumably due to their stylised  and thus recognisable presentation. (iii) This illustrates the value of a visual hashtag-centric approach: First, we gain the ability to recognise foods cross-culturally, and second, to recognise foods that may not have been prioritised by any top-down approach to create an image or food ontology. (iv) Somewhat surprisingly, there are multiple visual-hashtags that might appear to be abstract, but are actually stylised enough in presentation to be reliably recognisable, for example \#dessertporn, \#theartofplating, \#gastroart, and \#foodporno. (v) Constraining the domain in this way has a significant positive effect on the accuracy of the top-20 visual hashtags (75.1\% average accuracy in Fig.~\ref{tab:foodVisualPlate}, compared to 71.1\% in Figure~\ref{tab:foodVisual}. 

\begin{figure*}[t]
\centering
\includegraphics[width=0.3\textwidth]{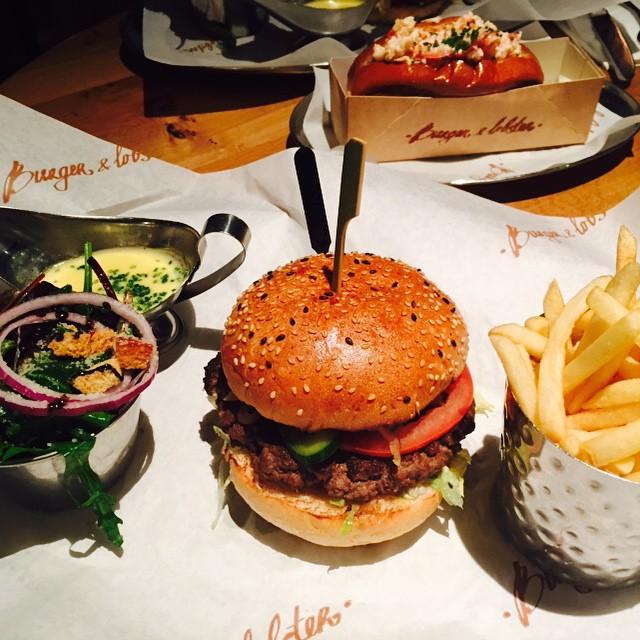}
\includegraphics[width=0.3\textwidth]{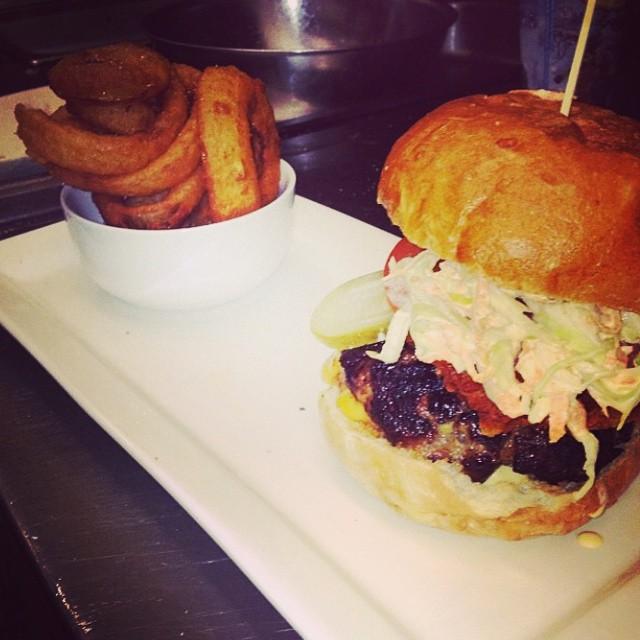}
\includegraphics[width=0.3\textwidth]{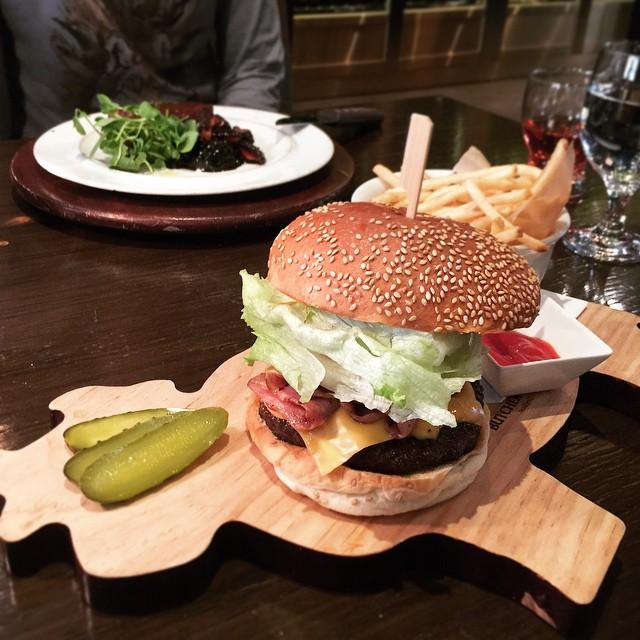}\\
\includegraphics[width=0.3\textwidth]{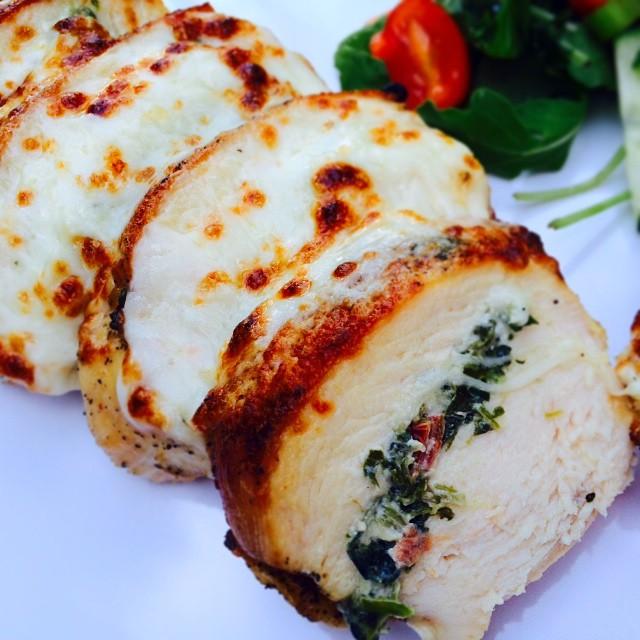}
\includegraphics[width=0.3\textwidth]{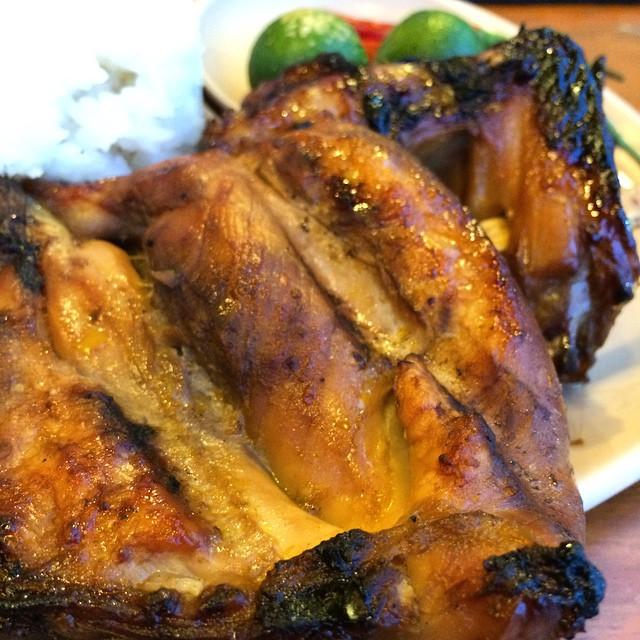}
\includegraphics[width=0.3\textwidth]{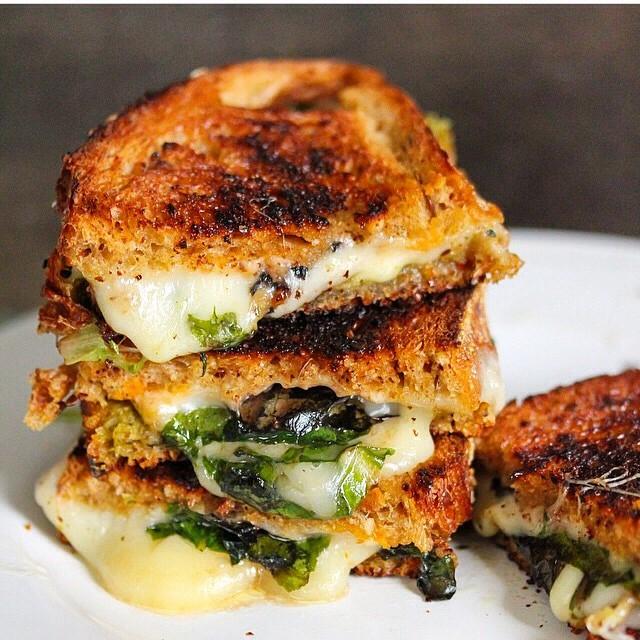}\\
\includegraphics[width=0.3\textwidth]{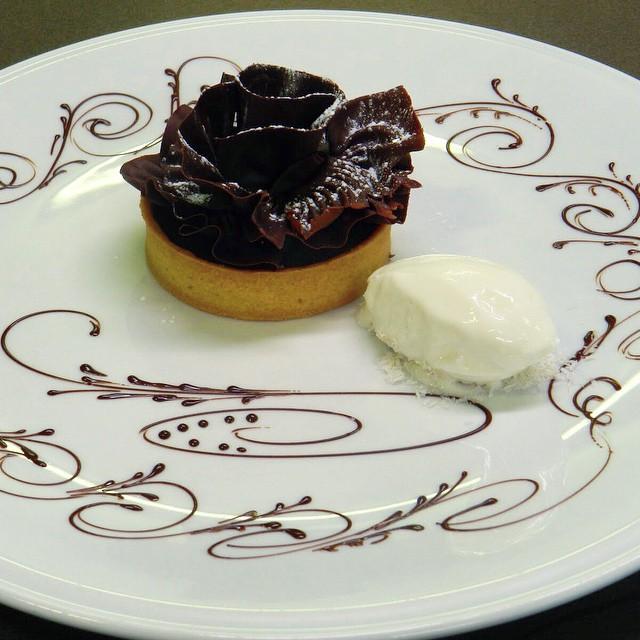}
\includegraphics[width=0.3\textwidth]{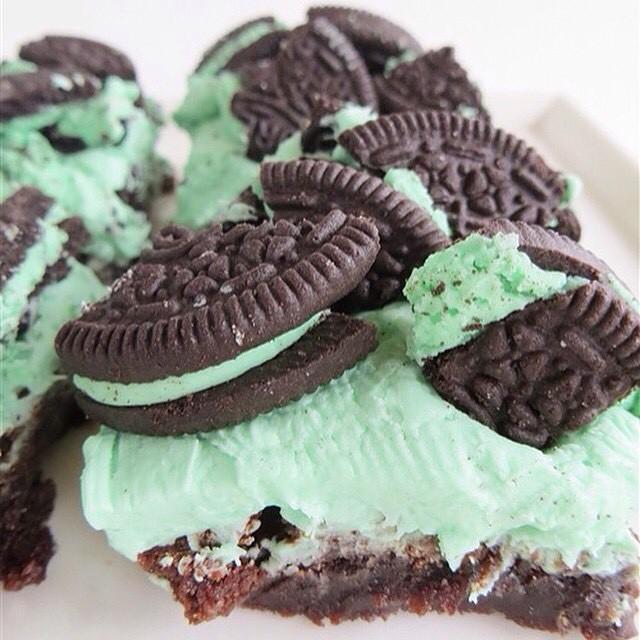}
\includegraphics[width=0.3\textwidth]{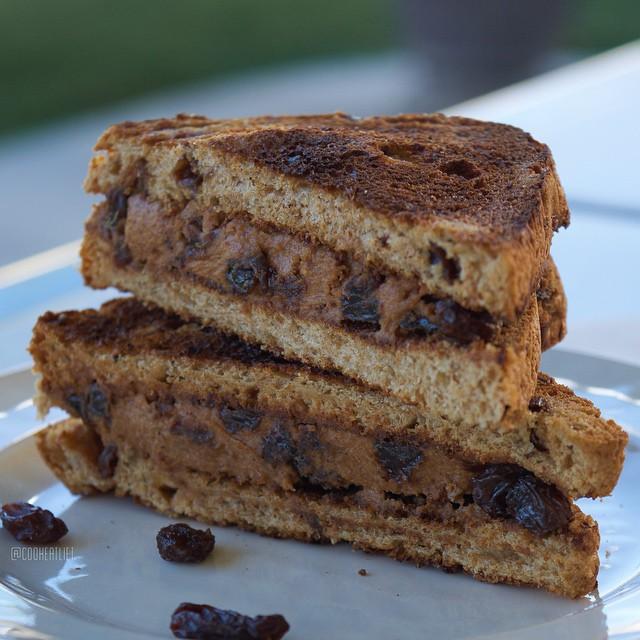}\\
\includegraphics[width=0.3\textwidth]{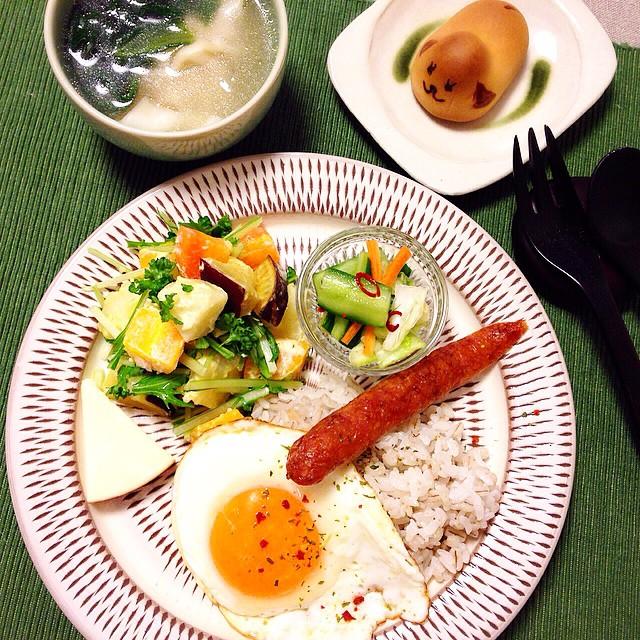}
\includegraphics[width=0.3\textwidth]{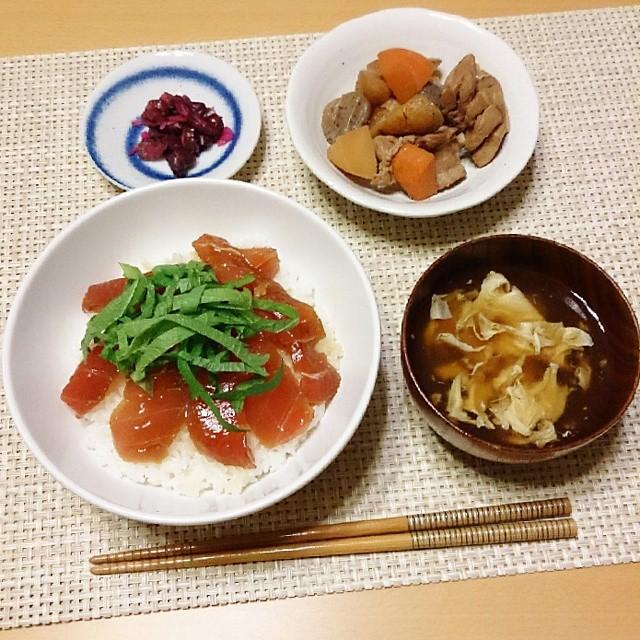}
\includegraphics[width=0.3\textwidth]{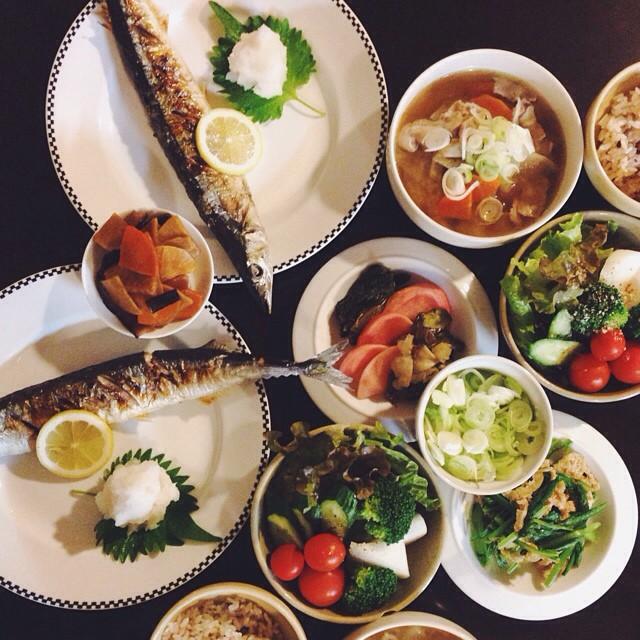}\\
\caption{Example images detected by our bottom-up learned visual concept recogniser. Top row: \#burger. Second row: \#foodporno, Third row: \#dessertporn. Bottom row: \#(japanese breakfast).}\label{fig:exampleImages}
\end{figure*}

Some qualitative examples of images confidently detected as matching particular hashtag categories are given in Figure~\ref{fig:exampleImages}. These examples cover categories well covered by traditional food ontologies such as hamburger (\#burger, Fig.~\ref{fig:exampleImages}, first row). More interestingly, we can see the visualisation of internet vernacular such as \#foodporno and \#dessertporn (Figure~\ref{fig:exampleImages}, second and third rows). From these we learn that \#foodporno typically refers to savoury foods in contrast to \#dessertporn's sweet foods, as well as the stylised photography of images warranting these tags -- which goes some way to explaining how an apparently tag concept can indeed be visually detectable in practice. Finally, we see examples images detected as matching  \#(Japanese Breakfast) tag (Figure~\ref{fig:exampleImages}, bottom row), illustrating the value of bottom-up learning of visual food categories from tags in covering categories that may have been missed by traditional top-down food ontologies.

\begin{table}[t]
\centering
\caption{(Raw). Top-20 most visual food hashtags. Rank: Overall visualness ranking. Freq Rank: Frequency of this hashtag in the dataset. Nrm Acc: Normalised (imbalance adjusted) classification accuracy is the visualness metric. Precision and Average Precision: Manual evaluation of resulting model.}
\resizebox{1.0\columnwidth}{!}{  
\begin{tabular}{llllll}
\hline
Rank & Hashtag   & Freq Rank & Nrm. Acc & Prec @ 50 & AP \\
\hline
10           & salad     & 43             & 74.2\%                    & 0.560          & 0.367                  \\
15           & smoothie  & 224            & 75.2\%                    & 0.640          & 0.473                  \\
16           & paella    & 325            & 76.5\%                    & 0.300          & 0.185                  \\
17           & cupcakes  & 306            & 71.8\%                    & 0.400          & 0.327                  \\
20           & chocolate & 34             & 72.3\%                    & 0.940          & 0.943                  \\
21           & burger    & 129            & 71.3\%                    & 0.280          & 0.267                  \\
22           & donut     & 277            & 70.5\%                    & 0.640          & 0.662                  \\
23           & pizza     & 97             & 70.7\%                    & 0.660          & 0.583                  \\
25           & dessert   & 32             & 68.6\%                    & 0.900          & 0.914                  \\
26           & ramen     & 451            & 77.0\%                    & 0.460          & 0.493                  \\
32           & cake      & 41             & 71.1\%                    & 0.620          & 0.644                  \\
35           & coffee    & 62             & 70.2\%                    & 0.600          & 0.611                  \\
36           & spaghetti & 226            & 71.3\%                    & 0.280          & 0.227                  \\
37           & soup      & 137            & 70.4\%                    & 0.040          & 0.039                  \\
40           & muffin    & 264            & 68.8\%                    & 0.160          & 0.140                  \\
41           & noodles   & 184            & 71.5\%                    & 0.240          & 0.235                  \\
42           & fries     & 193            & 66.2\%                    & 0.040          & 0.014                  \\
43           & cookie    & 228            & 66.2\%                    & 0.440          & 0.401                  \\
46           & avocado   & 239            & 69.4\%                    & 0.020          & 0.022                  \\
47           & juice     & 189            & 68.7\%                    & 0.160          & 0.144   \\              
\hline
\end{tabular}
}
\label{tab:foodVisual}
\end{table}

\begin{table}[t]
\centering
\caption{Results of Hashtag Learning (Plate Focus). Top-20 most visual food hashtags.}\label{tab:foodVisualPlate}

\resizebox{1.0\columnwidth}{!}{  
\begin{tabular}{llllll}
Rank                     & Hashtag                      & Freq Rank & Nrm Acc & Prec @ 20    & AP \\
1                        & breakfast (Japanese)       &                271                          & 83.8\%                   &         0.70          &        0.55                \\
2                        & breakfast (Traditional Japanese)                           & 343                          & 82.1\%                  &          0.325        &   0.317                 \\
3                        & home food (Japanese) 		& 253            & 77.1\%      &              0.525   &      0.54                  \\
4                        & fries                        & 140            & 71.4\%                    & 0.35              & 0.302                  \\
5                        & sushi                        & 100            & 75.4\%                    & 0.25              & 0.232                  \\
6                        & burger                       & 129            & 73.8\%                    & 0.35              & 0.373                  \\
7                        & dessertporn*                  & 218            & 74.7\%                    & 0.55             & 0.620                \\
8                        & roast                        & 179            & 70.6\%                    & 0.50              & 0.488                  \\
9                        & spaghetti                    & 163            & 75.2\%                    & 0.60              & 0.675                  \\
10                       & theartofplating*              & 295            & 79.2\%                    & 0.70             & 0.549                 \\
11                       & sweettooth                   & 220            & 74.4\%                    & 0.55              & 0.623                  \\
12                       & desserts                     & 187            & 73.4\%                    & 0.65              & 0.740                  \\
13                       & banana                       & 212            & 72.7\%                    & 0.20              & 0.240                  \\
14                       & pancakes                     & 272            & 73.1\%                    & 0.05              & 0.050                  \\
15                       & cake                         & 119            & 73.4\%                    & 0.45              & 0.665                  \\
16                       & gastroart*                   & 333            & 76.3\%                    & 0.90             & 0.925                 \\
17                       & soup                         & 170            & 69.0\%                    & 0.10              & 0.113                  \\
18                       & foodporno*                   & 238            & 77.3\%                    & 0.50             & 0.551                 \\
19                       & sashimi                      & 387            & 75.2\%                    & 0.35              & 0.444                  \\
20                       & icecream                     & 177            & 72.8\%                    & 0.05              & 0.066                 
\end{tabular}
}

\end{table}

\section{Further Analysis}
\subsection{Geospatial Distribution}

The popularity of different foods is broken down by continent in Tab~\ref{tab:vggFoodByContinent}. We present results for both VGG ImageNet-1000 food categories and our visual tag classifier (using those hashtags determined to be reliably visually recognisable in Sec~\ref{sec:visualHashtagResult}). Note that both of these results are based on visual analysis of the actual Instagram photos, in contrast to purely tag-based studies such as \cite{mejova2015foodporn}. 

From the top VGG classes, the most popular foods would appear to be relatively universal (ice cream, pizza are burrito are popular on every continent). However, there are also some unique aspects such as `Hot Pot' in Asia. From the visual hashtag analysis we see that \#sweettooth and \#dessertporn are now part of a global food vocabulary. It is interesting to note that North and South America document their beer consumption with greater frequency than their other continental counterparts. In addition, Africa is the only continent to have a savoury hashtag as their most popular. 

\begin{table*}[t]
\centering
\caption{Breakdown of Instagram posted foods by continent.}

\label{tab:vggFoodByContinent}
\begin{tabular}{clllllll}
\hline
& Overall   & Africa    & Asia      & Australia    & Europe          & N. America   & S. America      \\
\hline
\parbox[t]{2mm}{\multirow{5}{*}{\rotatebox[origin=c]{90}{VGG Cats.}}} 
& Ice cream & Pizza     & Ice cream & Ice cream    & Ice cream       & Burrito      & Ice cream       \\
& Burrito   & Burrito   & Carbonara & Cheeseburger & Pizza           & Ice cream    & Burrito         \\
& Pizza     & Bakery    & Hot pot   & Pizza        & Bakery          & Cheeseburger & Chocolate sauce \\
& Packet    & Ice cream & Burrito   & Burrito      & Packet          & Pizza        & Cheeseburger    \\
& Bakery    & Hotdog    & Bakery    & Hotdog       & Chocolate sauce & Hotdog       & Bakery         \\
\hline
\parbox[t]{2mm}{\multirow{11}{*}{\rotatebox[origin=c]{90}{Visual Hashtags}}} 
& Overall     & Africa      & Asia        & Australia   & Europe      & N. America  & S. America  \\
\hline
& Dessert     & Salad       & Dessert     & Dessert     & Dessert     & Dessert     & Chocolate   \\
& Chocolate   & Dessert     & Coffee      & Coffee      & Chocolate   & Salad       & Dessert     \\
& Cake        & Chocolate   & Cake        & Salad       & Cake        & Chocolate   & Cake        \\
& Salad       & Pizza       & Chocolate   & Chocolate   & Salad       & Coffee      & Salad       \\
& Coffee      & Coffee      & Salad       & Cake        & Desserts    & Cake        & Desserts    \\
& Desserts    & Cake        & Desserts    & Smoothie    & Dessertporn & Beer        & Dessertporn \\
& Sweettooth  & Sweettooth  & Sweettooth  & Dessertporn & Sweettooth  & Pizza       & Sweettooth  \\
& Dessertporn & Desserts    & Dessertporn & Pizza       & Coffee      & Desserts    & Coffee      \\
& Pizza       & Dessertporn & Pizza       & Foodporno   & Drink       & Sweettooth  & Beer        \\
& Drink       & Drink       & Drink       & Desserts    & Pizza       & Dessertporn & Drink      \\
\hline
\end{tabular}
\end{table*}

\begin{figure}[t]
\centering
\includegraphics[width=1.1\columnwidth]{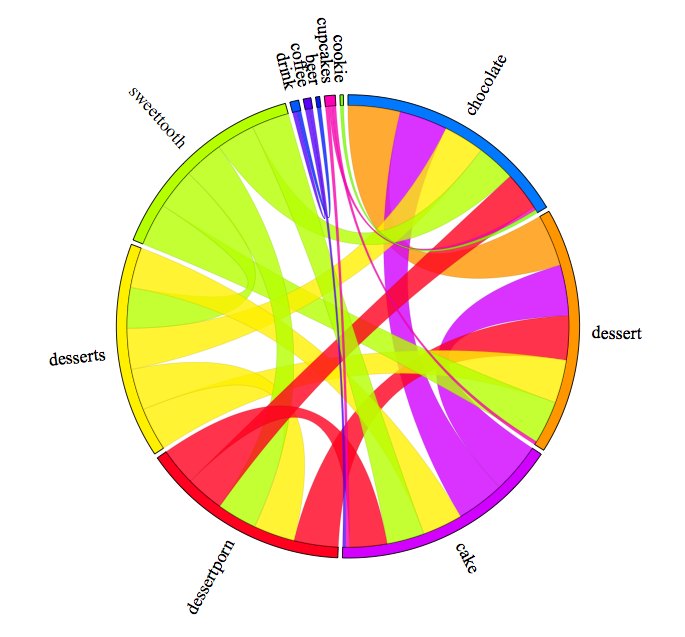}
\caption{Hashtag correlation across the dataset.}\label{fig:overallTagcorr}
\end{figure}

\subsection{Tag Correlations}

We also investigated the co-occurrence of visual categories across the dataset. Fig.~\ref{fig:overallTagcorr} visualises the strongest correlations in the resulting correlation matrix. This did not vary strongly across countries. Unsurprisingly, the greatest correlations were between the sweet items of the list. The hashtags \#sweettooth, \#desserts, \#dessert, \#chocolate, \#dessertporn, \#cake were the most strongly correlated. Since coffee is normally an accompaniment to dessert, positive correlations also existed between coffee and cake, and coffee and chocolate. These relationships were also analysed within each continent. All of the correlations were remarkably similar, with Europe and South America having the strongest correlations overall, and Australia having the weakest. 

The strongest negative correlations existed between \#salad and most of the other hashtags. While salad and the sweet hashtags had a negative relationship, salad also has negative correlations with coffee and drink. Additionally \#beer was infrequently accompanied by one of the dessert-related hashtags.

\section{Discussion}
\label{sec:disc}

Existing social media digital health studies \cite{mejova2015foodporn} have tended to focus on hashtags. This is a good start but suffers from the issue of only analysing what the user chooses to explicitly tag in an image. In contrast, analysing actually posted images has the possibility to reduce this bias to some degree\footnote{There is still bias according to which images are posted.} by analysing the actual image content, rather than what the user chose to tag about the image. Moreover, data can be obtained from all those un-tagged images if we can automatically understand their content.

However, despite some initial attempts~\cite{martin2009quantification, martin2014measuring,Sharma:2015:MCN:2740908.2742754}, analysing images automatically at scale is not easy, mainly due to: (i) the huge intra-class variability of many food types (e.g., salad composed of greens, grains, or fruit.); (ii) the variability of illumination, pose, image quality etc of photos taken by users with smartphone cameras; (iii) the requirement of typical recognition methods that the object to recognise should be centred and cropped of distracting background context; and (iv) the necessity of pre-defining food ontologies, as well as collecting annotated training data to train recognisers.

In this paper, we explored an alternative approaches to Instagram food image analysis: We first leveraged the deep neural network VGGnet. This is highly accurate and reliable, but only covers a small subset of 61 food categories included in its 1,000 category knowledge-base. Subsequently, we explored large scale training of classifiers to recognise individual hashtags, allowing consistently visual (rather than, e.g., emotional) hashtags to emerge bottom-up. This approach tended to be less accurate than VGGnet because the user-generated nature of the tags results in label-noise (false positive and false negative tagging) from the perspective of the classifier trying to learn a visual concept. However, it has the key benefits that (i) it has the potential to learn much wider and unbounded variety of food categories due to sidestepping the scalability barrier of collecting curated and annotated training data, and (ii) by learning from tags bottom-up, it avoids the need to pre-specify a top-down food ontology, and the associated cultural biases.

\paragraph{Lessons and Implications}

Although appealing, our approach of learning visual concepts bottom-up from tags faces some significant challenges. The central challenge is label noise: with false-positive (food-tagged but non-food content) as well as false negative (images missing some key content tags) being prevalent. In this paper we sketch some approaches that can help alleviate this, for example by filtering out non-food images and focusing on high-confidence food images to reduce noise level (Sec~\ref{sec:improve}). Going forward, more sophisticated approaches from the machine learning community that deal explicitly with label noise may be able to further address this issue \cite{reed2015cnnLabelnoise,sukhbaatar2015cnnLabelNoise}. A different challenge is introduced by our strategy of training classifiers for every tag, which is a very resource intensive exercise when performed at large scale, but is at least very amenable to parallelisation.  In future, false positive and missing tags could also be cleaned up by exploiting inter-tag-correlation (e.g., as we discovered, beer and dessert rarely co-occur)~\cite{deng2014objectClassifyLabelRelation}, however this may compromise the parallelism of the current framework.

Another factor contributing to label-noise was user incentives on Instagram. Some Instagram users aim to garner \emph{likes}, with some websites, such as ``Instagram Tags: More likes, More Followers" providing users with suggested hashtags, to accomplish just that, despite their possible irrelevance with relation to the image content. For instance, a selfie, or self-portrait, may contain the popular hashtag \#foodgasm, despite it being an unsuitable descriptor of the image, so that it may appear on the timeline of searches for that tag.

\section{Conclusions and Future Work}
In this paper we make a first attempt in large-scale, bottom-up, image content discovery using user-generated tags in Instagram image posts \emph{in-the-wild}. Our results are promising and show that in addition to identifying popular items such as sweets and despite the presence of noisy-labels, we can break the boundaries of traditional well-labeled training and testing machine-learning approaches, and use the hashtags to further discover categories which may not have been present in pre-labelled image categories. This approach is useful for analysing the users' interests and incentives when sharing food images, and consequently help in understanding individuals' perception of food that is visually appealing and the associated caloric values.

In future work we aim to investigate the ability of detecting various categories by relying purely on user labels for various tag types and compare the visualness of the tags when used in conjunction with different image types. Another avenue for potential future research is understanding the effect of presence of faces in images on the visualness and variety of tags used, hence comparing the social element versus culinary focus of images and their respective captions and comments.

%
%


\bibliographystyle{acm}
\bibliography{foodporn}

\begin{thebibliography}{10}

\bibitem{abbar2014you}
{\sc Abbar, S., Mejova, Y., and Weber, I.}
\newblock You tweet what you eat: Studying food consumption through twitter.
\newblock In {\em Proceedings of the 33rd Annual ACM Conference on Human
  Factors in Computing Systems\/} (New York, NY, USA, 2015), CHI '15, ACM,
  pp.~3197--3206.

\bibitem{berg2010automatic_web}
{\sc Berg, T.~L., Berg, A.~C., and Shih, J.}
\newblock Automatic attribute discovery and characterization from noisy web
  data.
\newblock In {\em ECCV\/} (Berlin, Heidelberg, 2010), ECCV'10, Springer-Verlag,
  pp.~663--676.

\bibitem{black1993measurements}
{\sc Black, A.~E., Prentice, A.~M., Goldberg, G.~R., Jebb, S.~A., Bingham,
  S.~A., Livingstone, M. B.~E., and Coward, A.}
\newblock Measurements of total energy expenditure provide insights into the
  validity of dietary measurements of energy intake.
\newblock {\em Journal of the American Dietetic Association 93}, 5 (1993),
  572--579.

\bibitem{chen2013neil}
{\sc Chen, X., Shrivastava, A., and Gupta, A.}
\newblock Neil: Extracting visual knowledge from web data.
\newblock In {\em ICCV\/} (December 2013).

\bibitem{deng2014objectClassifyLabelRelation}
{\sc Deng, J., Ding, N., Jia, Y., Frome, A., Murphy, K., Bengio, S., Li, Y.,
  Neven, H., and Adam, H.}
\newblock Large-scale object classification using label relation graphs.
\newblock In {\em ECCV}, D.~Fleet, T.~Pajdla, B.~Schiele, and T.~Tuytelaars,
  Eds., vol.~8689 of {\em Lecture Notes in Computer Science}. Springer
  International Publishing, 2014, pp.~48--64.

\bibitem{deng2009imagenet}
{\sc Deng, J., Dong, W., Socher, R., Li, L.-J., Li, K., and Fei-Fei, L.}
\newblock Imagenet: A large-scale hierarchical image database.
\newblock In {\em CVPR\/} (2009).

\bibitem{Duggan2015}
{\sc Duggan, M., Ellison, N.~B., Lampe, C., Lenhart, A., and Madden, M.}
\newblock Social media update 2014.
\newblock {\em Pew Research Center\/} (2015).

\bibitem{fan2008liblinear}
{\sc Fan, R.-E., Chang, K.-W., Hsieh, C.-J., Wang, X.-R., and Lin, C.-J.}
\newblock Liblinear: A library for large linear classification.
\newblock {\em Journal of Machine Learning Research 9\/} (June 2008),
  1871--1874.

\bibitem{morteza2016}
{\sc Fard, M.~A., Haddadi, H., and Targhi, A.~T.}
\newblock Fruits and vegetables calorie counter using convolutional neural
  networks.
\newblock In {\em ACM Digital Health\/} (April 2016).

\bibitem{fried2014analyzing}
{\sc Fried, D., Surdeanu, M., Kobourov, S., Hingle, M., and Bell, D.}
\newblock Analyzing the language of food on social media.
\newblock In {\em Big Data (Big Data), 2014 IEEE International Conference on\/}
  (2014), IEEE, pp.~778--783.

\bibitem{hu2014we}
{\sc Hu, Y., Manikonda, L., and Kambhampati, S.}
\newblock What we instagram: A first analysis of instagram photo content and
  user types.
\newblock In {\em ICWSM\/} (2014), AAAI.

\bibitem{Kawano:2014:FIR:2638728.2641339}
{\sc Kawano, Y., and Yanai, K.}
\newblock Food image recognition with deep convolutional features.
\newblock In {\em Proceedings of the 2014 ACM International Joint Conference on
  Pervasive and Ubiquitous Computing: Adjunct Publication\/} (New York, NY,
  USA, 2014), UbiComp '14 Adjunct, ACM, pp.~589--593.

\bibitem{layne2014wildAttr}
{\sc Layne, R., Hospedales, T., and Gong, S.}
\newblock Re-identification: Hunting attributes in the wild.
\newblock In {\em British Machine Vision Conference\/} (2014).

\bibitem{martin2009quantification}
{\sc Martin, C.~K., Kaya, S., and Gunturk, B.~K.}
\newblock Quantification of food intake using food image analysis.
\newblock In {\em Engineering in Medicine and Biology Society, 2009. EMBC 2009.
  Annual International Conference of the IEEE\/} (2009), IEEE, pp.~6869--6872.

\bibitem{martin2014measuring}
{\sc Martin, C.~K., Nicklas, T., Gunturk, B., Correa, J.~B., Allen, H., and
  Champagne, C.}
\newblock Measuring food intake with digital photography.
\newblock {\em Journal of Human Nutrition and Dietetics 27}, s1 (2014), 72--81.

\bibitem{mejova2015foodporn}
{\sc Mejova, Y., Haddadi, H., Noulas, A., and Weber, I.}
\newblock \# foodporn: Obesity patterns in culinary interactions.
\newblock {\em ACM conference on Digital Health 2015\/} (2015).

\bibitem{Oliva:2001:MSS:598425.598462}
{\sc Oliva, A., and Torralba, A.}
\newblock Modeling the shape of the scene: A holistic representation of the
  spatial envelope.
\newblock {\em Int. J. Comput. Vision 42}, 3 (May 2001), 145--175.

\bibitem{reed2015cnnLabelnoise}
{\sc Reed, S., Lee, H., Anguelov, D., Szegedy, C., Erhan, D., and Rabinovich,
  A.}
\newblock Training deep neural networks on noisy labels with bootstrapping.
\newblock In {\em ICLR Workshop\/} (2015).

\bibitem{Sharma:2015:MCN:2740908.2742754}
{\sc Sharma, S.~S., and De~Choudhury, M.}
\newblock Measuring and characterizing nutritional information of food and
  ingestion content in instagram.
\newblock In {\em Proceedings of the 24th International Conference on World
  Wide Web\/} (Republic and Canton of Geneva, Switzerland, 2015), WWW '15
  Companion, International World Wide Web Conferences Steering Committee,
  pp.~115--116.

\bibitem{simonyan2015veryDeep}
{\sc Simonyan, K., and Zisserman, A.}
\newblock Very deep convolutional networks for large-scale image recognition.
\newblock In {\em ICLR\/} (2015).

\bibitem{sukhbaatar2015cnnLabelNoise}
{\sc Sukhbaatar, S., Bruna, J., Paluri, M., Bourdev, L., and Fergus, R.}
\newblock Training convolutional networks with noisy labels.
\newblock In {\em ICLR Workshop\/} (2015).

\bibitem{zepeda2008think}
{\sc Zepeda, L., and Deal, D.}
\newblock Think before you eat: photographic food diaries as intervention tools
  to change dietary decision making and attitudes.
\newblock {\em International Journal of Consumer Studies 32}, 6 (2008),
  692--698.

\end{thebibliography}

\end{document}